\documentclass[aps,floatfix,prb,reprint,notitle,
,superscriptaddress]{revtex4-2}
\usepackage{graphicx} 
\usepackage{chemformula}
\usepackage{float}
\usepackage{comment}
\usepackage{appendix}
\usepackage{siunitx}

\begin{document}

\preprint{APS/123-QED}
\title{Hydrogen bond symmetrization in high-pressure ice clathrates}

\author{Lorenzo Monacelli}
 \affiliation{Dipartimento di Fisica, Universit\`a di Roma Sapienza}
 
\author{Maria Rescigno}
 \affiliation{Dipartimento di Fisica, Universit\`a di Roma Sapienza, Piazzale Aldo Moro 5, 00185 Roma, Italy}
 \affiliation{Institut Laue-Langevin, 71 Avenue des Martyrs, Cedex 9, Grenoble, France}
 
\author{Alasdair Nicholls}
 \affiliation{Laboratory of Quantum Magnetism, Institute of Physics, \'{E}cole Polytechnique F\'{e}d\'{e}erale de Lausanne, CH-1015 Lausanne, Switzerland}
 
\author{Umbertoluca Ranieri}
 \affiliation{Centre for Science at Extreme Conditions and School of Physics and Astronomy, University of Edinburgh, EH9 3FD Edinburgh, UK}
 \affiliation{Centro de Física de Materiales (CFM-MPC), CSIC-UPV/EHU, Donostia/San Sebastián 20018, Spain}
 
\author{Simone Di Cataldo}
 \affiliation{Dipartimento di Fisica, Universit\`a di Roma Sapienza, Piazzale Aldo Moro 5, 00185 Roma, Italy}
 \affiliation{Instit\"{u}t f\"{u}r Festk\"{o}rperphysik, TU Wien, Wiedner Hauptstraße 8-10, Vienna 1050, Austria}
 
\author{Livia Eleonora Bove}
 \affiliation{Dipartimento di Fisica, Universit\`a di Roma Sapienza, Piazzale Aldo Moro 5, 00185 Roma, Italy}
 \affiliation{Laboratory of Quantum Magnetism, Institute of Physics, \'{E}cole Polytechnique F\'{e}d\'{e}erale de Lausanne, CH-1015 Lausanne, Switzerland}
 \affiliation{Institut de Min\'{e}ralogie, de Physique des Mat\'{e}riaux et de Cosmochimie (IMPMC), Sorbonne Universit\'{e}, CNRS UMR 7590,  MNHN, 4, place Jussieu, Paris, France}

\begin{abstract}
Hydrogen bond symmetrization is a fundamental pressure-induced transformation in which the distinction between donor and acceptor sites vanishes, resulting in a symmetric hydrogen-bond network. While extensively studied in pure ice—most notably during the ice VII to ice X transition—this phenomenon remains less well characterized in hydrogen hydrates. In this work, we investigate hydrogen bond symmetrization in the high-pressure phases of hydrogen hydrate (\ch{H2-H2O} and \ch{H2-D2O}) through a combined approach of Raman spectroscopy and first-principles quantum atomistic simulations. We focus on the C2 and C3 filled-ice phases, using both hydrogenated and deuterated water frameworks. Our results reveal that quantum fluctuations and the interaction between the encaged \ch{H2} molecules and the host lattice play a crucial role in driving the symmetrization process. Remarkably, we find that in both C2 and C3 phases, hydrogen bond symmetrization occurs via a continuous crossover at significantly lower pressures than in pure ice, without any change in the overall crystal symmetry. These findings provide new insight into the quantum-driven mechanisms of bond symmetrization in complex hydrogen-bonded systems under extreme conditions.
\end{abstract}

\maketitle

\section{Introduction}
Hydrogen bond symmetrization (HBS) is a fundamental phenomenon in hydrogen-rich materials, profoundly altering their structural, vibrational, elastic, and electronic properties under extreme conditions. It has been extensively studied in pure water ice \cite{Aoki96, Gonchy1996, pruzan_raman_1997, Komatsu2023}, where it drives the transition from hydrogen-disordered ice VII to the symmetric hydrogen-bonded ice X phase \cite{cherubini_quantum_2024}. Similar transitions have been identified in other hydrogen-bonded systems. In salt-filled ice structures, both experimental and computational studies have shown that ionic species can induce symmetrization of hydrogen bonds by compressing the water network and modifying the proton potential energy landscape \cite{Bove2015, Bronstein2016}. Methane-filled ice phases also exhibit HBS, with Raman evidence indicating that symmetrization governs the transformation from the MH-III to the MH-IV phase \cite{Schaack2018, Schaack2019}. In hydrous minerals such as $\delta$-AlOOH, high-pressure neutron diffraction has revealed a universal onset of symmetrization at a critical O···O distance of 2.443(1) Å \cite{meier2022structural}. Despite these advances, HBS in hydrogen hydrates remains largely unexplored \cite{ranieri_observation_2023, dicataldo2024}, even though these materials provide an ideal setting to study the interplay between hydrogen bonding and guest–host interactions under pressure.

Hydrogen hydrate (HH) is known to crystallize in five distinct phases: the low-temperature clathrate structure sII, and four high-pressure filled-ice or framework phases labeled C0, C1, C2, and C3. The sII phase consists of a water framework encapsulating hydrogen molecules within large polyhedral cages \cite{Lokshin2004, ranieri_quantum_2019, ranieri_large-cage_2024}. C0, the first high-pressure phase, is a chiral gas-hydrate structure distinct from conventional ice polymorphs, and is stable between 0.35 and 0.7 GPa \cite{Amos2017, Strobel2016, delRosso2016}. C1, stable from 0.9 to 2.7 GPa (and metastable up to 5.2 GPa) \cite{kuzovnikov2019, Vos1996, Carvalho2021}, adopts an ice II framework, while C2 and C3 are based on an expanded ice Ic lattice with 1:1 and 1:2 water-to-hydrogen ratios, respectively \cite{Lokshin2004, ranieri_observation_2023}. The C2 phase becomes stable above 2.7 GPa and transforms into C3 above 30 GPa at ambient temperature. C3 remains stable at least up to 90 GPa and metastable down to 7 GPa \cite{Leon2025}.

Among these, the C2 phase exhibits notable structural similarities to ice VII. Its water framework mirrors one of the two interpenetrating diamond sublattices of ice VII, while the second sublattice is occupied by \ch{H2} molecules. The result is an expanded cubic ice lattice \cite{komatsu_2020}, offering a compelling model to explore HBS under the competing influences of quantum effects and guest–host interactions. The recently discovered C3 phase \cite{ranieri_observation_2023}, built on the same water lattice but incorporating twice as many \ch{H2} molecules, enables an investigation of symmetrization in an even denser hydrogen environment. These systems are not only of fundamental interest—shedding light on how symmetrization occurs in an expanded and dynamically perturbed ice lattice—but also have implications for understanding high-pressure processes in hydrogen-rich planetary interiors, where such materials are likely to exist \cite{Leon2025, Tobie24}.

While symmetrization in ice VII has been observed experimentally \cite{Aoki96, Gonchy1996, pruzan_raman_1997, Komatsu2023} and studied through quantum simulations \cite{cherubini_quantum_2024, Benoit2002, Razvan_2018}, discrepancies remain regarding the precise transition pressure, reflecting the challenges of locating hydrogen atoms under extreme conditions. In hydrogen hydrates, these challenges are compounded by the presence of the \ch{H2} sublattice, which introduces additional interactions and dynamical effects. Quantum tunneling, already central in pure ice, may play an even more prominent role here due to the low mass and high mobility of the \ch{H2} molecules.

In this work, we extend the study of hydrogen bond symmetrization to the C2 and C3 phases of hydrogen hydrate. Using quantum atomistic simulations and Raman spectroscopy, we identify the onset of symmetrization and characterize its structural and vibrational signatures. We find that in both phases, symmetrization occurs through a smooth crossover, without modifying the crystal symmetry. In the C2 phase, in particular, HBS occurs at a much lower pressure than in pure ice. These results highlight the key role of quantum fluctuations and guest–host coupling in shaping HBS in complex hydrogen-bonded frameworks.

\section{Results and Discussion}

Hydrogen bond symmetrization in ice can be probed experimentally through Raman spectroscopy, where the \ch{OH} stretching mode exhibits softening, broadening, and loss of intensity under increasing pressure. In ice VII, this mode becomes undistinguishable from background in the spectra before reaching full symmetrization as a result of severe broadening and intensity reduction. In pure ice, the appearance of a \textit{T}\textsubscript{2g} mode, indicative of the highly symmetric cubic ice X lattice, serves as an alternative signature of the transition.

\begin{figure*}[hbtp]
\centering
\includegraphics[width=1\columnwidth]{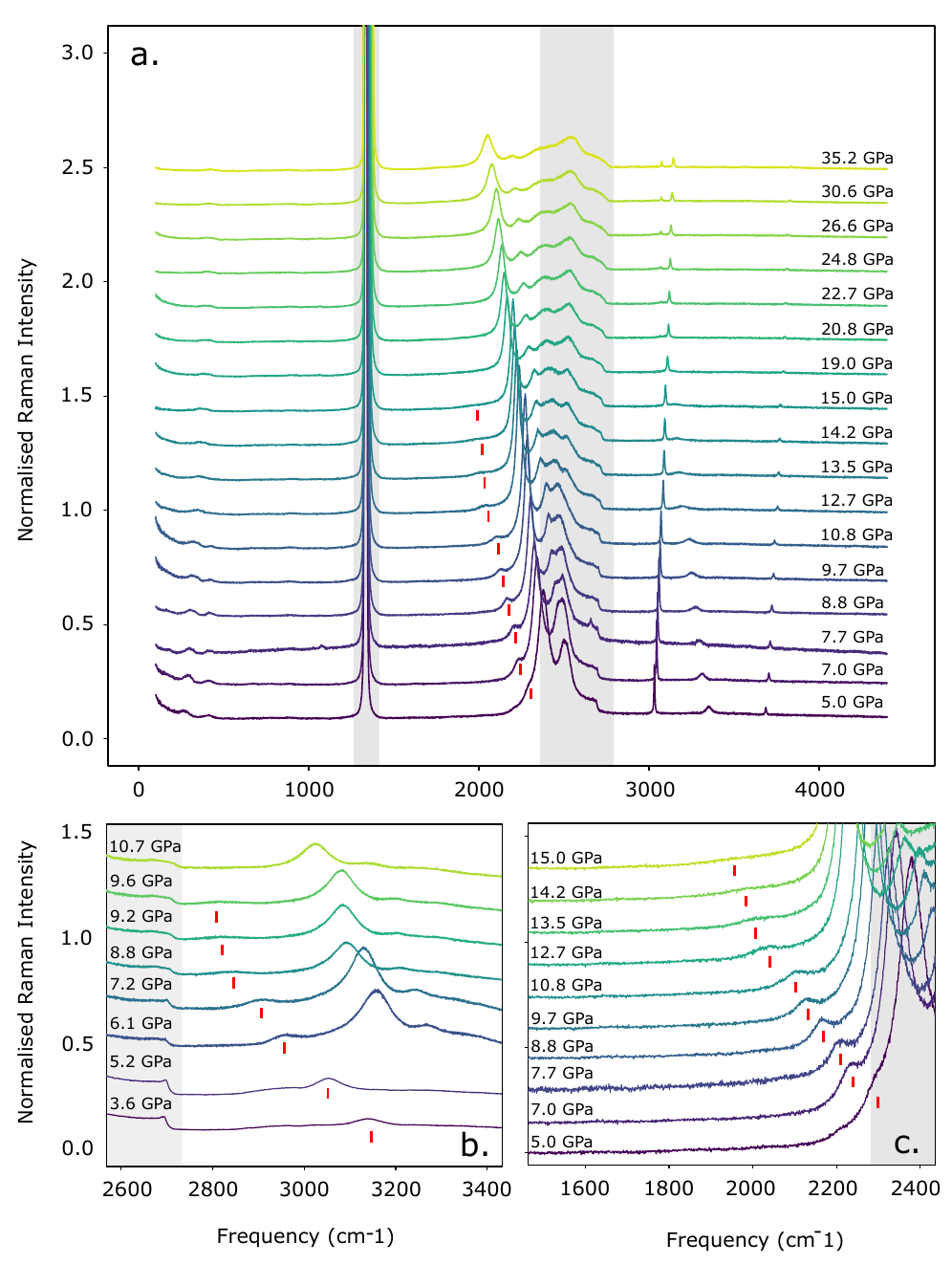}
\caption{Raman spectra measured in the \ch{D2O}-\ch{D2} (top and bottom right panel) and \ch{H2O}-\ch{H2} (bottom left panel) filled ice C2 structure as a function of pressure. The red tick indicates the position of hydrate stretching mode. The shaded area indicates the region where the diamond Raman activity covers the sample signal.}
\label{fig:Raman:spectrum}
\end{figure*}

Raman spectra for hydrogen-filled ice C2 at selected pressures are shown in \figurename~\ref{fig:Raman:spectrum}, for both \ch{H2:D2O} (top and bottom left panels) and \ch{H2:H2O} (bottom right panel). The hydrate stretching mode, marked by a red tick, undergoes a progressive redshift and broadening with pressure, reflecting the weakening of the donor–acceptor distinction in the hydrogen bond. A shaded region indicates the spectral range in which diamond anvil Raman activity obscures the sample signal. A higher frequency peak corresponds to the OD/OH stretching mode of ice VII, formed during the C1 to C2 phase transition \cite{ranieri_observation_2023}. The isotopic comparison reveals distinct pressure-dependent shifts and dispersions for \ch{D2O} and \ch{H2O}, enabling a direct evaluation of the effects of quantum mass.

Compared to ice VII, the OH-stretching mode in the C2 phase softens more strongly, suggesting that hydrogen-bond symmetrization occurs at lower pressures. However, signal attenuation, peak broadening, and overlap with the second-order diamond band above \SI{10}{\giga\pascal} limit its spectroscopic visibility. No clear secondary Raman features emerge to indicate symmetrization directly. To complement these limitations, we performed first-principles simulations.

We used the Stochastic Self-Consistent Harmonic Approximation (SSCHA) \cite{monacelli_stochastic_2021, miotto_fast_2024, monacelli_black_2020, monacelli_quantum_2023} to account for quantum fluctuations and the anharmonicity of the light atoms involved. SSCHA approximates the quantum ionic wavefunction as a Gaussian that is optimized to minimize the free energy. The method has been successfully applied to describe hydrogen symmetrization in pure ice \cite{cherubini_quantum_2024}. The interatomic forces were modeled with an equivariant neural network potential based on NequIP \cite{batzner_e3-equivariant_2022}, trained on DFT data from previous studies \cite{ranieri_observation_2023,dicataldo2024}. The simulations were performed using a proton-ordered model of the ice sublattice, which contains two symmetry-inequivalent \ch{H2O} molecules in the primitive cell, as in ice VIII and XI.

\begin{figure*}[hbtp]
\centering
\includegraphics[width=\linewidth]{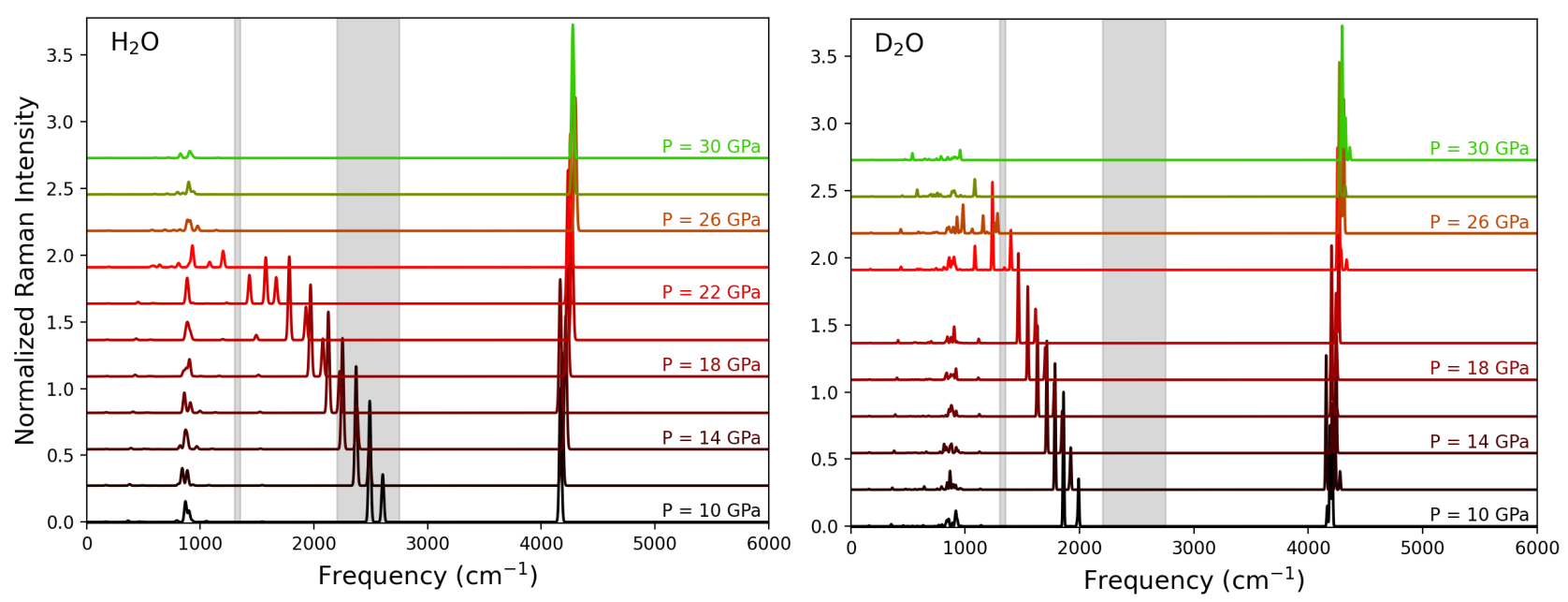}
\caption{Anharmonic Raman spectrum computed within the SCHA in the \ch{H2O} (left panel) and \ch{D2O} (right panel) \ch{H2} filled ice C2 structure. The shaded area indicates the region where the diamond Raman activity covers the sample signal.}
\label{fig:SCHA:spectrum}
\end{figure*}

The computed anharmonic Raman spectra for \ch{H2O} and \ch{D2O} C2 are shown in \figurename~\ref{fig:SCHA:spectrum}. A qualitative change in the spectral response is observed between \SI{24}{\giga\pascal} and \SI{26}{\giga\pascal}, corresponding to the disappearance of all hydrate-related stretching modes.

The simulations reveal three main Raman-active modes before symmetrization between 1000~cm$^{-1}$ and 3000~cm$^{-1}$: two from mixed symmetric and antisymmetric stretching, and a third bending mode near \SI{1500}{\per\centi\meter} that becomes enhanced via mode coupling around \SI{22}{\giga\pascal}. After symmetrization, all three disappear, consistent with the experiment. The \ch{H2} molecule remains spectroscopically silent in this range, as it does not participate in lattice vibrations relevant to the transition, progressively hardening above \SI{4000}{\per\centi\meter} upon compression. The small peak slightly below \SI{1000}{\per\centi\meter} represents the \ch{H2} rotations, which, however, are not accurately described within the SSCHA framework\cite{siciliano_beyond_2024,siciliano_beyond_2024-1}, and require a different treatment to be accounted for\cite{dicataldo2024}.

These spectra are derived from SSCHA auxiliary phonons, which account for anharmonic frequency renormalization but not finite phonon lifetimes \cite{bianco_second-order_2017, monacelli_time-dependent_2021, siciliano_wigner_2023}, thereby explaining the absence of the pronounced broadening observed in experiments. However, this method enables accurate mode tracking with pressure.

\begin{figure}[hbtp]
\centering
\includegraphics[width=\linewidth]{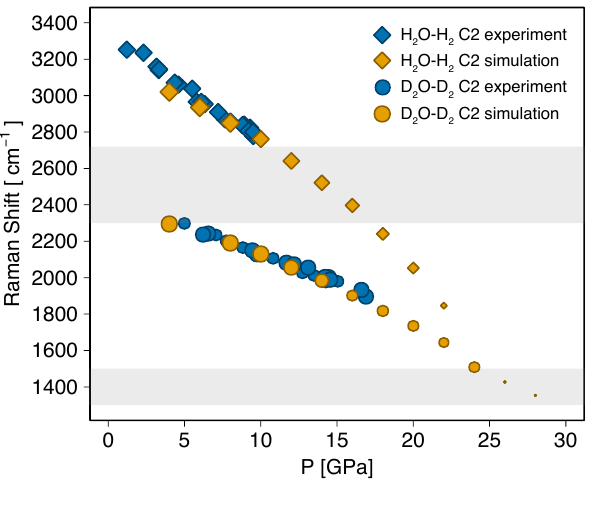}
\caption{Comparison of experimental (blue) and calculated (orange) frequencies of the OH (losange) and OD (dots) stretching modes as a function of pressure. The shaded area indicates the region where the diamond Raman activity covers the sample signal. Symbol sizes of simulation data are proportional to intensities and only the higher intensity peak is reported. In Figure S3 of the Supplementary Information, we show the intensity behavior of the experimental data. A global rigid shift in frequencies of \SI{270}{\per\centi\meter} is applied to simulation data to correct the known mismatch of semi-local DFT functionals on the OH stretching frequency of ice\cite{Cherubini2021}.}
\label{fig:frequencies:comparison}
\end{figure}

\figurename~\ref{fig:frequencies:comparison} compares the simulated and experimental stretching frequencies for \ch{H2O} and \ch{D2O}. A rigid frequency shift of \SI{270}{\per\centi\meter} corrects known DFT underestimations of the OH stretching~\cite{Cherubini2021}. Experimental data show a redshift consistent with bond strengthening and eventual mode disappearance above \SI{12}{\giga\pascal} for \ch{H2O} and \SI{17}{\giga\pascal} for \ch{D2O}. Despite these limitations, the pressure dependence is well reproduced, validating the simulations.

\begin{figure}[hbtp]
\centering
\includegraphics[width=\linewidth]{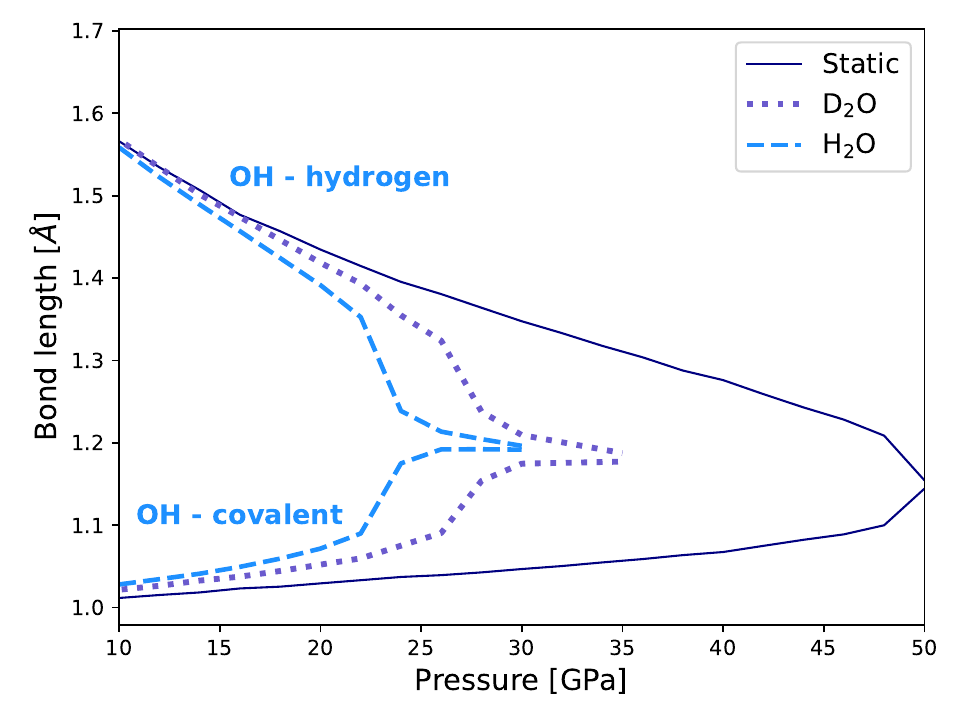}
\caption{OH bonds as a function of pressure. Comparison of static (no quantum or thermal fluctuations), deuterium (\ch{D2O}), and protium (\ch{H2O}) symmetrization of the OH bond. Inequivalent molecules in the proton-ordered model we employed have been averaged.}
\label{fig:structure}
\end{figure}

Despite the spectral changes near \SI{24}{\giga\pascal}, no analogue of the ice X \textit{T}\textsubscript{2g} mode is observed. This absence can be attributed to structural differences: unlike ice VII/VIII, which consists of two interpenetrating hydrogen-bonded networks, the C2 structure replaces one with \ch{H2}, reducing volume and stabilizing symmetrization at lower pressures. 
The decreasing Raman intensity and broadening prevent tracking the stretching mode down to \SI{0}{\per\centi\meter}, a hallmark of the phase transition. For this reason, we computed the average ionic positions from the quantum ionic wavefunction and evaluated the bond lengths. As shown in \figurename~\ref{fig:structure}, the OH bond sharply changes between \SI{22}{\giga\pascal} and \SI{25}{\giga\pascal}, coinciding with the loss of Raman activity in the simulations. In contrast to ice VIII-X\cite{cherubini_quantum_2024}, symmetrization in this system occurs gradually. 

The isotope effect is evident: replacing \ch{H} with \ch{D} shifts the crossover by approximately \SI{5}{\giga\pascal}, as shown in \figurename~\ref{fig:structure}b. This pressure shift and the suppression of symmetrization in static calculations reflect strong quantum nuclear effects in hydrogen bonding, a hallmark of ice physics.

\begin{figure}[H]
\centering
\includegraphics[width=\linewidth]{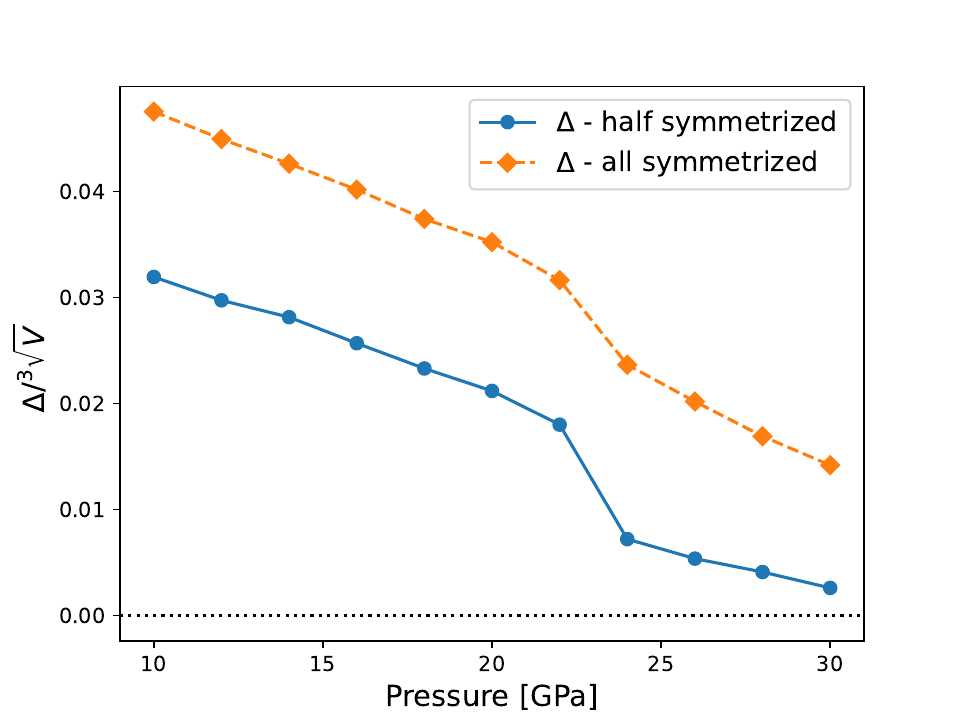}
\caption{Distance of the ice clathrate to the configurations where half of the covalent bonds are symmetrized, and all covalent bonds are symmetrized. The symmetry group considers the proton-ordered version of the ice lattice. The 3rd root of the volume rescales the distance to a volume-independent, dimensionless parameter.}
\label{fig:struct:delta}
\end{figure}

To shed light on the crossover character of the hydrogen bond symmetrization in hydrates, we performed a symmetry analysis removing the \ch{H2} guest molecule. The symmetrization of the two independent sublattices of the proton-ordered model we employed is slightly delayed, indicating that, even in the proton-disordered phase, not all the molecules symmetrize simultaneously. Moreover, unlike ice VII-X\cite{cherubini_quantum_2024}, one sublattice symmetrization is more pronounced. By defining the order parameter $\Delta$ as the difference between the distance of the hydrogen atom from the near neighbour oxygens ($\Delta = d_{OH} - d_{HO}$), we can see the continuous character of the phase transition plotting $\Delta$ as a function of pressure, in \figurename~\ref{fig:struct:delta}. A sharp change of $\Delta$ toward partial symmetrization occurs between \SI{22}{\giga\pascal} and \SI{24}{\giga\pascal}, coinciding with the loss of the Raman signal. This behavior arises from the presence of two subgroups with slightly different O–H bond lengths in the proton-ordered structures. Full symmetrization proceeds more gradually than in empty ice, clearly marking a gradual crossover rather than a sharp phase transition. This behavior contrasts with the second-order transition in pure ice near \SI{55}{\giga\pascal} between phases VIII and X \cite{cherubini_quantum_2024}, highlighting the interactions between the water network and the host molecular hydrogen.

\begin{figure*}[hbtp]
    \centering
    \includegraphics[width=\linewidth]{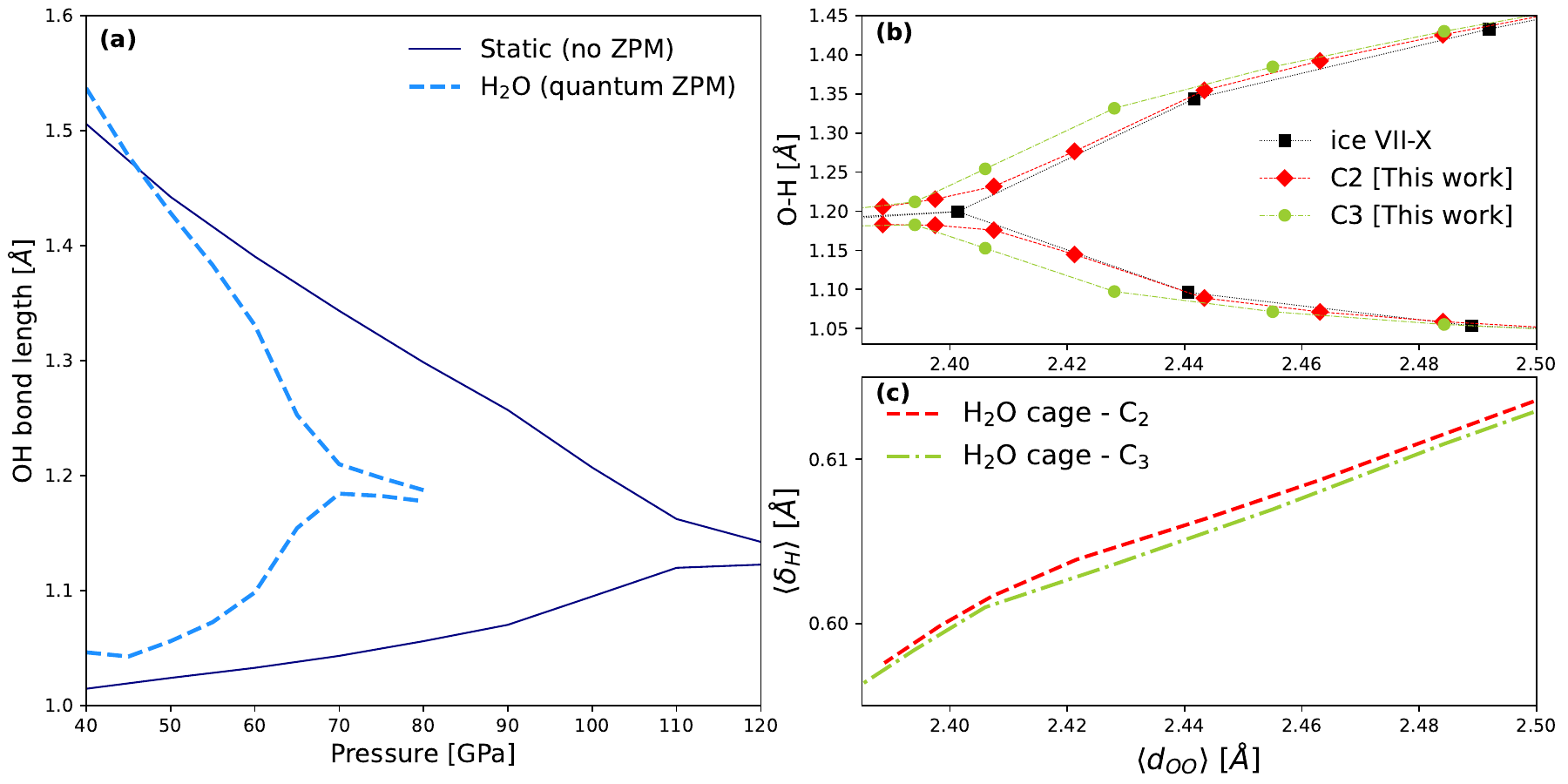}
    \caption{\textbf{(a)} OH bond symmetrization in the ice sublattice in the C3 phase. Comparison between the simulation accounting for nuclear quantum fluctuations (dashed light blue line) with the static result neglecting quantum ionic fluctuations (solid navy line). \textbf{(b)} Comparison of the OH bond symmetrization vs O-O distance in different systems: pure ice VII\cite{cherubini_quantum_2024}, C2 and C3. While the C2 and pure ICE symmetrize at the same O-O distance, the C3 requires a smaller O-O distance to symmetrize, shedding light on the role played by the environment in hydrogen symmetrization. \textbf{(c)} Quantum average displacement of the H-bonded atom perpendicular to the oxygen-oxygen distance vector. The C3 perpendicular displacement of H atoms is inhibited by the presence of dense \ch{H2} molecules for the same size of the \ch{H2O} cage compared to the C2 phase.}
    \label{fig:universal}
\end{figure*}

We compare the symmetrization of hydrogen in the C2 phase with that observed in pure ice and the high-pressure C3 structure. In the C3 phase, the host \ch{H2O} lattice is structurally equivalent to that of C2, but the molar ratio between \ch{H2} and \ch{H2O} is 2:1—the highest ever reported\cite{ranieri_observation_2023}.
Despite the structural similarity of the ice cages in C2 and C3, the stretching mode in C3 is strongly mixed with low-frequency molecular translational modes. This results in negligible Raman activity for the mode associated with symmetrization, making the detection of hydrogen symmetrization via vibrational spectroscopy experimentally unfeasible.

Symmetrization in C3 occurs at significantly higher pressures than in C2 due to the larger volume of the ice sublattice, caused by the greater number of hydrogen molecules confined within the ice channels, as shown in \figurename~\ref{fig:universal}\textbf{(a)}. Also in this case, quantum fluctuations play a crucial role, lowering the transition pressure from \SI{110}{\giga\pascal} to \SI{70}{\giga\pascal}. As in C2, the symmetrization proceeds through a smooth crossover rather than a sharp phase transition.

To analyze the environmental influence on hydrogen bond symmetrization, we plot the O–H bond length as a function of the average O–O distance (\figurename~\ref{fig:universal}\textbf{(b)}). In both C2 and pure ice\cite{cherubini_quantum_2024}, symmetrization occurs at an O–O distance of approximately \SI{2.41}{\angstrom}. In contrast, C3 symmetrizes at a shorter O–O distance of \SI{2.39}{\angstrom}, despite sharing the same ice sublattice. This deviation underscores the impact of the local environment: the denser population of \ch{H2} molecules in C3 reduces out-of-plane hydrogen fluctuations within the ice lattice, thereby requiring a shorter O–O distance to achieve symmetrization.

A universal picture of hydrogen bond symmetrization emerges when we examine the OH bond as a function of O–O distance in pure ice, C2, and the C3 phase. While C2 and pure ice symmetrize at the same critical geometry, C3 requires a shorter O–O distance by \SI{0.01}{\angstrom}, despite sharing the same water framework (\figurename~\ref{fig:universal}\textbf{c}). This deviation highlights the importance of the local environment: in C3, the dense hydrogen guest network suppresses the transverse quantum fluctuations of hydrogen atoms, as shown in \figurename~\ref{fig:universal}\textbf{c}, thereby delaying symmetrization and altering its spectroscopic fingerprint. In particular, the OH stretching is almost completely suppressed at such high pressures in the C3, preventing the usual spectroscopic identification of the hydrogen symmetrization.

\section{Conclusions}

We have investigated hydrogen bond symmetrization in the hydrogen-filled ice C2 phase and compared it with both pure ice and the higher-density C3 phase. Despite the structural similarity of the water frameworks in C2 and C3, the spectroscopic signatures of symmetrization differ substantially.

In the C2 phase, Raman spectroscopy, supported by quantum anharmonic simulations, reveals a clear softening and eventual disappearance of the OH stretching mode, consistent with a continuous symmetrization process occurring around \SI{24}{\giga\pascal}. The transition is gradual and strongly influenced by quantum nuclear effects, as shown by the isotope-dependent shift and the inability to induce symmetrization in static calculations. This behavior contrasts with the well-defined second-order transition observed in pure ice near \SI{55}{\giga\pascal}, but occurs at the same critical O–O distance of approximately \SI{2.41}{\angstrom}.

In the C3 phase, however, vibrational spectroscopy fails to detect symmetrization due to strong mixing of the OH stretching mode with low-frequency translational modes. This suppresses Raman activity, making the transition spectroscopically silent. Nonetheless, simulations confirm that symmetrization does occur -- at significantly higher pressures than in C2 -- due to the larger volume of the ice lattice caused by the higher hydrogen loading (2:1 \ch{H2}:\ch{H2O}). Quantum effects again reduce the transition pressure dramatically, from \SI{110}{\giga\pascal} to \SI{70}{\giga\pascal}, but even here the process remains a smooth crossover rather than a sharp phase change.

These findings underscore the intricate interplay between lattice geometry, guest molecule density, and quantum nuclear effects in controlling hydrogen bond symmetrization. Our results demonstrate that structural similarity alone does not determine the symmetrization mechanism: the host–guest interaction and quantum dynamics must be considered to fully understand and predict the behavior of hydrogen-bonded systems under pressure.

\section*{Acknowledgments}
L.M. acknowledges the CINECA award under the ISCRA initiative for the availability of high-performance computing resources and support.

\section*{Appendix: Numerical methods}

    The SSCHA performs a Gaussian ansatz on the ionic density matrix $\hat \rho$, which depends on the average nuclear position $\boldsymbol{\mathcal R}$ and the covariance matrix of quantum/thermal fluctuations. This covariance matrix is defined through an auxiliary force constant matrix $\boldsymbol{\Phi}$ which, at convergence, satisfies the self-consistent relation:
    \begin{equation}
        \Phi_{ab} = \left< \frac{d^2V}{dR_adR_b}\right>_{\hat\rho(\boldsymbol{\mathcal R}, \boldsymbol{\Phi})},
    \end{equation}
    where $\left<\circ\right>_{\hat\rho(\boldsymbol{\mathcal R}, \boldsymbol{\Phi})}$ indicates the average on an ensemble extracted according the Gaussian density matrix and $V(\boldsymbol{R})$ is the Born-Oppenheimer energy landscape with the nuclei fixed in positions $\boldsymbol{R}$. We can diagonalize the auxiliary dynamical matrix and extract eigenvalues and eigenvectors of phonons as
    \begin{equation}
        \sum_b \frac{\Phi_{ab}e_\mu^b}{\sqrt{m_a m_b}} = \omega_\mu^2 e_\mu^a 
    \end{equation}
    where $\omega_\mu$ are the anharmonic dressed phonon frequencies, and $e_\mu$ is the associated polarization vector. The Raman intensity of phonons is evaluated by contracting the respective polarization vector with the Raman tensor $\boldsymbol{\Xi}$:
    $$
    \Xi_{abc} = \frac{d^3 V}{dE_a dE_b dR_c} 
    $$
    \begin{equation}
        I_\mu = \frac{1 + n_\mu}{\omega_\mu} \sum_{abc}\Xi_{abc}e_\mu^c \epsilon_a \epsilon_b
    \end{equation}
    where $\boldsymbol{\epsilon}$ is the polarization of the light. More details are reported in refs\cite{monacelli_stochastic_2021,miotto_fast_2024}. 
    The real anharmonic frequencies should be evaluated from the time-dependent response function\cite{monacelli_time-dependent_2021}; however, the anharmonic dressed frequencies $\omega_\mu$ already account for anharmonicity and are a good approximation far from phase transitions, even in strongly anharmonic crystals\cite{monacelli_simulating_2024}. To obtain the average structure, we performed a variable cell relaxation at fixed pressure $P$ by minimizing the Gibbs free energy:
    $$
    G(\boldsymbol{\Phi}, \boldsymbol{\mathcal R}, \{\boldsymbol{a_i}\}) = \left<\hat H\right>_{\hat\rho(\boldsymbol{\mathcal R}, \boldsymbol{\Phi})} - TS[\boldsymbol{\Phi}] + P\Omega(\{\boldsymbol{a_i}\})
    $$
    where $\hat H$ is the Born-Oppenheimer Hamiltonian, $S$ is the entropy (which only depends on $\boldsymbol{\Phi}$), and $\Omega$ is the volume, function of the lattice parameters $\{\boldsymbol{a_i}\}$. Further details on the algorithm to relax the primitive cell are discussed in ref.\cite{monacelli_pressure_2018}.

    The Born-Oppenheimer energy landscape $V(\boldsymbol{R})$ is approximated via an equivariant machine-learning force field based on NequIP\cite{batzner_e3-equivariant_2022}. The model is trained on a set of 800 first-principles calculations on a 2x2x1 supercell of the C2 and C3 phase randomly selected from equilibrium SSCHA relaxation performed within density functional theory, as performed in a previous work\cite{ranieri_observation_2023} (see there for all the details on the DFT calculation). The data are sampled from pressures uniformly distributed through \SI{5}{\giga\pascal} and \SI{40}{\giga\pascal} of the C2 phase and from \SI{40}{\giga\pascal} and \SI{80}{\giga\pascal} of the C3 phase. The dataset is divided into 650 configurations for training and 150 for testing. The root mean squared error on the energy in the test set is \SI{1.3}{\milli\electronvolt} per atom and \SI{90.9}{\milli\electronvolt/\angstrom} for oxygen forces and \SI{58.7}{\milli\electronvolt/\angstrom} for hydrogen forces. We benchmark the full calculation with the trained potential with the result obtained at \SI{20}{\giga\pascal} and \SI{30}{\giga\pascal} for DFT in a previous calculation and found that all the relevant observables (OH bond length, equation of state, anharmonic vibrational frequencies) are well converged and reproduce consistently the DFT results.

\clearpage


\begin{thebibliography}{10}

\bibitem{Aoki96}
K.~Aoki, H.~Yamawaki, M.~Sakashita, and H.~Fujihisa, ``Infrared absorption
  study of the hydrogen-bond symmetrization in ice to 110 gpa,'' {\em Phys.
  Rev. B}, vol.~54, pp.~15673--15677, Dec 1996.

\bibitem{Gonchy1996}
A.~F. Goncharov, V.~V. Struzhkin, M.~S. Somayazulu, R.~J. Hemley, and H.~K.
  Mao, ``Compression of ice to 210 gigapascals: Infrared evidence for a
  symmetric hydrogen-bonded phase,'' {\em Science}, vol.~273, no.~5272,
  pp.~218--220, 1996.

\bibitem{pruzan_raman_1997}
P.~Pruzan, E.~Wolanin, M.~Gauthier, J.~C. Chervin, B.~Canny, D.~Häusermann,
  and M.~Hanfland, ``Raman {Scattering} and {X}-ray {Diffraction} of {Ice} in
  the {Megabar} {Range}. {Occurrence} of a {Symmetric} {Disordered} {Solid}
  above 62 {GPa},'' {\em The Journal of Physical Chemistry B}, vol.~101,
  pp.~6230--6233, Aug. 1997.
\newblock Publisher: American Chemical Society.

\bibitem{Komatsu2023}
K.~Komatsu, T.~Hattori, S.~Klotz, S.~Machida, K.~Yamashita, H.~Ito,
  H.~Kobayashi, T.~Irifune, T.~Shinmei, A.~Sano-Furukawa, and H.~Kagi,
  ``Hydrogen bond symmetrisation in {D2O} ice observed by neutron
  diffraction,'' {\em Nature Communications}, vol.~15, p.~5100, June 2024.

\bibitem{cherubini_quantum_2024}
M.~Cherubini, L.~Monacelli, B.~Yang, R.~Car, M.~Casula, and F.~Mauri, ``Quantum
  effects in {H}-bond symmetrization and in thermodynamic properties of high
  pressure ice,'' {\em Physical Review B}, vol.~110, p.~014112, July 2024.
\newblock Publisher: American Physical Society.

\bibitem{Bove2015}
L.~E. Bove, R.~Gaal, Z.~Raza, A.-A. Ludl, S.~Klotz, A.~M. Saitta, A.~F.
  Goncharov, and P.~Gillet, ``Effect of salt on the h-bond symmetrization in
  ice,'' {\em Proceedings of the National Academy of Sciences}, vol.~112,
  pp.~8216--8220, 7 2015.

\bibitem{Bronstein2016}
Y.~Bronstein, P.~Depondt, L.~E. Bove, R.~Gaal, A.~M. Saitta, and F.~Finocchi,
  ``Quantum versus classical protons in pure and salty ice under pressure,''
  {\em Physical Review B}, vol.~93, p.~024104, 1 2016.

\bibitem{Schaack2018}
S.~Schaack, U.~Ranieri, P.~Depondt, R.~Gaal, W.~F. Kuhs, A.~Falenty, P.~Gillet,
  F.~Finocchi, and L.~E. Bove, ``Orientational {Ordering}, {Locking}-in, and
  {Distortion} of {CH4} {Molecules} in {Methane} {Hydrate} {III} under {High}
  {Pressure},'' {\em The Journal of Physical Chemistry C}, vol.~122,
  pp.~11159--11166, May 2018.
\newblock Publisher: American Chemical Society.

\bibitem{Schaack2019}
S.~Schaack, U.~Ranieri, P.~Depondt, R.~Gaal, W.~F. Kuhs, P.~Gillet,
  F.~Finocchi, and L.~E. Bove, ``Observation of methane filled hexagonal ice
  stable up to 150 gpa,'' {\em Proceedings of the National Academy of
  Sciences}, vol.~116, no.~33, pp.~16204--16209, 2019.

\bibitem{meier2022structural}
T.~Meier, F.~Trybel, S.~Khandarkhaeva, D.~Laniel, T.~Ishii, A.~Aslandukova,
  N.~Dubrovinskaia, and L.~Dubrovinsky, ``Structural independence of
  hydrogen-bond symmetrisation dynamics at extreme pressure conditions,'' {\em
  Nature Communications}, vol.~13, no.~1, p.~3042, 2022.

\bibitem{ranieri_observation_2023}
U.~Ranieri, S.~Di~Cataldo, M.~Rescigno, L.~Monacelli, R.~Gaal, M.~Santoro,
  L.~Andriambariarijaona, P.~Parisiades, C.~De~Michele, and L.~E. Bove,
  ``Observation of the most {H2}-dense filled ice under high pressure,'' {\em
  Proceedings of the National Academy of Sciences}, vol.~120, p.~e2312665120,
  Dec. 2023.
\newblock Publisher: Proceedings of the National Academy of Sciences.

\bibitem{dicataldo2024}
S.~Di~Cataldo, M.~Rescigno, L.~Monacelli, U.~Ranieri, R.~Gaal, S.~Klotz,
  J.~Ollivier, M.~M. Koza, C.~De~Michele, and L.~E. Bove, ``Giant splitting of
  the hydrogen rotational eigenenergies in the c2 filled ice,'' {\em Phys. Rev.
  Lett.}, vol.~133, p.~236101, Dec 2024.

\bibitem{Lokshin2004}
K.~A. Lokshin, Y.~Zhao, D.~He, W.~L. Mao, H.-K. Mao, R.~J. Hemley, M.~V.
  Lobanov, and M.~Greenblatt, ``Structure and dynamics of hydrogen molecules in
  the novel clathrate hydrate by high pressure neutron diffraction,'' {\em
  Physical Review Letters}, vol.~93, p.~125503, 9 2004.

\bibitem{ranieri_quantum_2019}
U.~Ranieri, M.~M. Koza, W.~F. Kuhs, R.~Gaal, S.~Klotz, A.~Falenty,
  D.~Wallacher, J.~Ollivier, P.~Gillet, and L.~E. Bove, ``Quantum {Dynamics} of
  {H2} and {D2} {Confined} in {Hydrate} {Structures} as a {Function} of
  {Pressure} and {Temperature},'' {\em The Journal of Physical Chemistry C},
  vol.~123, pp.~1888--1903, Jan. 2019.
\newblock Publisher: American Chemical Society.

\bibitem{ranieri_large-cage_2024}
U.~Ranieri, L.~del Rosso, L.~E. Bove, M.~Celli, D.~Colognesi, R.~Gaal, T.~C.
  Hansen, M.~M. Koza, and L.~Ulivi, ``Large-cage occupation and quantum
  dynamics of hydrogen molecules in {sII} clathrate hydrates,'' {\em The
  Journal of Chemical Physics}, vol.~160, p.~164706, Apr. 2024.

\bibitem{Amos2017}
D.~M. Amos, M.-E. Donnelly, P.~Teeratchanan, C.~L. Bull, A.~Falenty, W.~F.
  Kuhs, A.~Hermann, and J.~S. Loveday, ``A chiral gas–hydrate structure
  common to the carbon dioxide–water and hydrogen–water systems,'' {\em The
  Journal of Physical Chemistry Letters}, vol.~8, pp.~4295--4299, 9 2017.

\bibitem{Strobel2016}
T.~A. Strobel, M.~Somayazulu, S.~V. Sinogeikin, P.~Dera, and R.~J. Hemley,
  ``Hydrogen-stuffed, quartz-like water ice,'' {\em Journal of the American
  Chemical Society}, vol.~138, no.~42, pp.~13786--13789, 2016.
\newblock PMID: 27540626.

\bibitem{delRosso2016}
L.~del Rosso, F.~Grazzi, M.~Celli, D.~Colognesi, V.~Garcia-Sakai, and L.~Ulivi,
  ``Refined structure of metastable ice xvii from neutron diffraction
  measurements,'' {\em The Journal of Physical Chemistry C}, vol.~120, no.~47,
  pp.~26955--26959, 2016.

\bibitem{kuzovnikov2019}
M.~Kuzovnikov and M.~Tkacz, ``T--p--x phase diagram of the water--hydrogen
  system at pressures up to 10 kbar,'' {\em The Journal of Physical Chemistry
  C}, vol.~123, no.~6, pp.~3696--3702, 2019.

\bibitem{Vos1996}
W.~L. Vos, L.~W. Finger, R.~J. Hemley, and H.-k. Mao, ``Novel
  ${\mathrm{h}}_{2}$-${\mathrm{h}}_{2}$o clathrates at high pressures,'' {\em
  Phys. Rev. Lett.}, vol.~71, pp.~3150--3153, Nov 1993.

\bibitem{Carvalho2021}
P.~H.~B. Brant~Carvalho, A.~Mace, I.~M. Nangoi, A.~A. Leitão, C.~A. Tulk,
  J.~J. Molaison, O.~Andersson, A.~P. Lyubartsev, and U.~Häussermann,
  ``Exploring high-pressure transformations in low-z (h2, ne) hydrates at low
  temperatures,'' {\em Crystals}, vol.~12, no.~1, 2022.

\bibitem{Leon2025}
L.~Andriambariarijaona, T.~Poreba, S.~D. Cataldo, R.~Gaal, U.~Ranieri,
  M.~Santoro, T.~Hansen, G.~Tobie, and L.~E. Bove, ``Hydrogen hydrates under
  extreme conditions: Insights into high-pressure phases and implications for
  planetary interiors,'' {\em Physical Review B}, in press.

\bibitem{komatsu_2020}
K.~Komatsu, S.~Machida, F.~Noritake, T.~Hattori, A.~Sano-Furukawa, R.~Yamane,
  K.~Yamashita, and H.~Kagi, ``Ice {Ic} without stacking disorder by evacuating
  hydrogen from hydrogen hydrate,'' {\em Nature Communications}, vol.~11,
  p.~464, Feb. 2020.

\bibitem{Tobie24}
T.~Van~Hoolst, G.~Tobie, C.~Vallat, N.~Altobelli, L.~Bruzzone, H.~Cao,
  D.~Dirkx, A.~Genova, H.~Hussmann, L.~Iess, J.~Kimura, K.~Khurana,
  A.~Lucchetti, G.~Mitri, W.~Moore, J.~Saur, A.~Stark, A.~Vorburger,
  M.~Wieczorek, A.~Aboudan, J.~Bergman, F.~Bovolo, D.~Breuer, P.~Cappuccio,
  L.~Carrer, B.~Cecconi, G.~Choblet, F.~De~Marchi, M.~Fayolle, A.~Fienga,
  Y.~Futaana, E.~Hauber, W.~Kofman, A.~Kumamoto, V.~Lainey, P.~Molyneux,
  O.~Mousis, J.~Plaut, W.~Puccio, K.~Retherford, L.~Roth, B.~Seignovert,
  G.~Steinbrügge, S.~Thakur, P.~Tortora, F.~Tosi, M.~Zannoni, S.~Barabash,
  M.~Dougherty, R.~Gladstone, L.~I. Gurvits, P.~Hartogh, P.~Palumbo, F.~Poulet,
  J.-E. Wahlund, O.~Grasset, and O.~Witasse, ``Geophysical {Characterization}
  of the {Interiors} of {Ganymede}, {Callisto} and {Europa} by {ESA}’s
  {JUpiter} {ICy} moons {Explorer},'' {\em Space Science Reviews}, vol.~220,
  p.~54, July 2024.

\bibitem{Benoit2002}
M.~Benoit, A.~H. Romero, and D.~Marx, ``Reassigning hydrogen-bond centering in
  dense ice,'' {\em Physical Review Letters}, vol.~89, p.~145501, 9 2002.

\bibitem{Razvan_2018}
C.~R. Hernandez~JA, ``Proton dynamics and the phase diagram of dense water
  ice.,'' {\em The Journal of Chemical Physics}, vol.~148, p.~214501, June
  2018.
\newblock Publisher: American Chemical Society.

\bibitem{monacelli_stochastic_2021}
L.~Monacelli, R.~Bianco, M.~Cherubini, M.~Calandra, I.~Errea, and F.~Mauri,
  ``The stochastic self-consistent harmonic approximation: calculating
  vibrational properties of materials with full quantum and anharmonic
  effects,'' {\em Journal of Physics: Condensed Matter}, vol.~33, p.~363001,
  July 2021.
\newblock Publisher: IOP Publishing.

\bibitem{miotto_fast_2024}
M.~Miotto and L.~Monacelli, ``Fast prediction of anharmonic vibrational spectra
  for complex organic molecules,'' {\em npj Computational Materials}, vol.~10,
  pp.~1--9, Oct. 2024.
\newblock Publisher: Nature Publishing Group.

\bibitem{monacelli_black_2020}
L.~Monacelli, I.~Errea, M.~Calandra, and F.~Mauri, ``Black metal hydrogen above
  360 {GPa} driven by proton quantum fluctuations,'' {\em Nature Physics},
  vol.~17, pp.~63--67, Sept. 2020.
\newblock Publisher: Springer Science and Business Media LLC.

\bibitem{monacelli_quantum_2023}
L.~Monacelli, M.~Casula, K.~Nakano, S.~Sorella, and F.~Mauri, ``Quantum phase
  diagram of high-pressure hydrogen,'' {\em Nature Physics}, vol.~19,
  pp.~845--850, June 2023.
\newblock Publisher: Nature Publishing Group.

\bibitem{batzner_e3-equivariant_2022}
S.~Batzner, A.~Musaelian, L.~Sun, M.~Geiger, J.~P. Mailoa, M.~Kornbluth,
  N.~Molinari, T.~E. Smidt, and B.~Kozinsky, ``E(3)-equivariant graph neural
  networks for data-efficient and accurate interatomic potentials,'' {\em
  Nature Communications}, vol.~13, p.~2453, May 2022.
\newblock Number: 1 Publisher: Nature Publishing Group.

\bibitem{siciliano_beyond_2024}
A.~Siciliano, L.~Monacelli, and F.~Mauri, ``Beyond {Gaussian} fluctuations of
  quantum anharmonic nuclei,'' {\em Physical Review B}, vol.~110, p.~134111,
  Oct. 2024.
\newblock Publisher: American Physical Society.

\bibitem{siciliano_beyond_2024-1}
A.~Siciliano, L.~Monacelli, and F.~Mauri, ``Beyond {Gaussian} fluctuations of
  quantum anharmonic nuclei: {The} case of rotational degrees of freedom,''
  {\em Physical Review B}, vol.~110, p.~144101, Oct. 2024.
\newblock Publisher: American Physical Society.

\bibitem{bianco_second-order_2017}
R.~Bianco, I.~Errea, L.~Paulatto, M.~Calandra, and F.~Mauri, ``Second-order
  structural phase transitions, free energy curvature, and
  temperature-dependent anharmonic phonons in the self-consistent harmonic
  approximation: {Theory} and stochastic implementation,'' {\em Physical Review
  B}, vol.~96, p.~014111, July 2017.
\newblock Publisher: American Physical Society (APS).

\bibitem{monacelli_time-dependent_2021}
L.~Monacelli and F.~Mauri, ``Time-dependent self-consistent harmonic
  approximation: {Anharmonic} nuclear quantum dynamics and time correlation
  functions,'' {\em Physical Review B}, vol.~103, p.~104305, Mar. 2021.
\newblock Publisher: American Physical Society (APS).

\bibitem{siciliano_wigner_2023}
A.~Siciliano, L.~Monacelli, G.~Caldarelli, and F.~Mauri, ``Wigner {Gaussian}
  dynamics: {Simulating} the anharmonic and quantum ionic motion,'' {\em
  Physical Review B}, vol.~107, p.~174307, May 2023.
\newblock Publisher: American Physical Society.

\bibitem{Cherubini2021}
M.~Cherubini, L.~Monacelli, and F.~Mauri, ``The microscopic origin of the
  anomalous isotopic properties of ice relies on the strong quantum anharmonic
  regime of atomic vibration,'' {\em Journal of Chemical Physics}, vol.~155,
  p.~184502, Nov. 2021.
\newblock Publisher: AIP Publishing.

\bibitem{monacelli_simulating_2024}
L.~Monacelli, ``Simulating anharmonic crystals: {Lights} and shadows of
  first-principles approaches,'' {\em ArXiV}, vol.~2407.03090, July 2024.
\newblock arXiv:2407.03090.

\bibitem{monacelli_pressure_2018}
L.~Monacelli, I.~Errea, M.~Calandra, and F.~Mauri, ``Pressure and stress tensor
  of complex anharmonic crystals within the stochastic self-consistent harmonic
  approximation,'' {\em Physical Review B}, vol.~98, p.~024106, July 2018.
\newblock Publisher: American Physical Society (APS).

\end{thebibliography}
\end{document}